\documentclass[%
 reprint,
 amsmath,amssymb,
 aps,pre
]{revtex4-2}

\usepackage{graphicx}% Include figure files
\usepackage{dcolumn}% Align table columns on decimal point
\usepackage{bm}% bold math
\usepackage{amsmath}
\usepackage{amssymb}
\usepackage{graphicx}
\usepackage{epstopdf}
\usepackage{epsfig}
\usepackage{physics}
\usepackage{hyperref}
\usepackage{lipsum}
\usepackage{graphics}
\usepackage{xcolor}
\usepackage{bbold}
\usepackage{mathtools}

\begin{document}

\preprint{APS/123-QED}

\title{The few-body problem of point particles with only gravitational interactions}

\author{Deepak Dhar}
\email{deepak@iiserpune.ac.in}
\affiliation{Indian Institute of Science Education and Research, Dr. Homi Bhabha Road, Pashan,  Pune 411008, India}

\date{\today}
\begin{abstract}
In this article, I discuss the motion of  $N$  point masses in nonrelativistic mechanics, when the interaction between them is purely the Newtonian gravitational interaction, with $N\ge 2$.  The dynamical  equations of motion cannot be solved in closed form, for general initial conditions, for any $N >2$. However, the qualitative behavior of the solutions can be understood from general considerations.   I discuss in particular the motion of three masses on a line, and the counterinituitive case of four masses on  a line  that can lead to all particles escaping to infinity in  finite time. 
\end{abstract}

\maketitle
% \onecolumngrid

\section{Introduction}

Kepler's laws, describing the motion of planets around the sun, were proposed based on extensive  astronomical observations.   Newton showed that these laws can be derived by assuming  that the  planet  moves around the sun under a force that varies as inverse square of the instantaneous distance from the sun.  Newton's derivation of the elliptical planetary orbits is a masterpiece of immense beauty.  If I was asked to identify two key problems  from the  undergraduate physics curriculum,  that best demonstrate the beauty and power of the Physicist's Way of looking at the Universe, one would be this, and the other is the quantum mechanical treatment of the energy levels of the  hydrogen atom. 

Given the importance of the Kepler problem in fostering the scientific world-view in students (agents like Rahu and Ketu are not invoked), it is a bit disappointing that it is not easy to find some simple generalizations, or extensions of this result in introductory Physics courses.  Of course, it is  natural to ask is  what happens if there are more than two gravitationally interacting masses. This question has puzzled  generations of physicists, including  some of the best minds,  like Lagrange,  and Poincare.  However, this subject,  called Few-body problems, is usually discussed in  rather technical  and highly mathematical papers and books, somewhat off-putting  to beginning students. My aim in this article is to discuss this subject, at a level that a good undergraduate can follow easily.  It is hoped that the article will serve  as an appetizer for  the more inquisitive students.

\section{Preliminaries}

We will discuss the time evolution of a system of $N$  point masses, in three dimensional space, with only the  gravitational interaction between them. We will discuss this in the Newtonian mechanics setting, where the kinetic energy is quadratic in the momentum, ignoring relativistic effects, and the gravitational force  is  proportional to the inverse  square of the instantaneous distance between the masses. 

If $\vec{r}_i$ is the position of the $i$-th mass at time $t$, the equations of motion are

\begin{equation}
\frac{d^2}{dt^2} \vec{r}_i =  -G \sum_{j \neq i}  \frac{m_j (\vec{r}_i - \vec{r}_j)}{|\vec{r}_i -\vec{r}_j|^3}
\end{equation}

The general question is to determine $\vec{r}_i$ as  functions of time, or at least determine their qualitative behavior. For example,  the question of the stablity of the Sun-Jupiter-Earth system   has been investigated much, but is still unsettled.

It is easy to see  that total momentum of the system, the total energy and the angular momentum about the center of mass are constants of motion. However, even for the $N=3$ case, this only gives $13$ separable variables, out of  $18$, and the problem cannot be solved in closed form, as posed here. 

Some qualitative aspects of the long-time behavior of gravitating systems are easy to guess, from our experience with other mechanical systems. For example, one can expect that, in general, at long times, the system will break into smaller clusters, where the particles within a cluster are bound gravitationally, so that their mutual distances remain finite, but the distances between centers of mass of different clusters increase linearly with time. This is true in general. But in 1971, D. G. Saari showed that there are very special initial conditions, under which 5 particles, moving solely under the gravitational field of each other go off to infinity in  finite time! This is rather unexpected, and may even sound impossible at first. This is what we will try to discuss in this article. An accessible account of earlier work on this topic, and additional references   may be found in \cite{bisbal, saari, hut}. 

\section{The scaling solutions}

We start with the study of some  special cases where the problem is more tractable, and the results are instructive. Firstly, if all the positions and velocities are in a plane, or on a line, then the subsequent motion remains on the  same plane, or line, respectively. Restricting the motion to a line, or a plane is a great simplification.

One simple case is where each particle executes a linear motion in the center of mass frame, and the configuration at one time is related to that at any other time, by simple  time-dependent scale factor. 
Let us  look for a solution of the form 
\begin{equation}
\vec{r}_i = R(t)  \vec{a}_i,
\end{equation}
where $\vec{a}_i$ are some constant vectors, satisfying $\sum_i  m_i \vec{a}_i =0$. Then, one finds that 
$R(t)$ and $\{a_i\}$  must satisfy the equations
\begin{eqnarray}
\frac{d^2}{dt^2} R(t) = -G \lambda / R(t)^2,\\
\vec{a}_i =  (\frac{1}{\lambda} )  \sum_{j \neq i}   \frac{ \vec{a}_i -\vec{a}_j}{| \vec{a}_i -\vec{a}_j|^3}.
\end{eqnarray}

Here the first equation  is easily solved for $R(t)$ ( see below). It is easily seen that $R(t)$ eventually starts decraesing, and goes to zero at a finite time $T^*$, when there is an $N$-body collision. For the time close to  this singularity, say for  $t = T^* - \Delta T$, $R(t)$ varies as $(\Delta T)^{2/3}$.

Equations (4)  are finite polynomial equations in the euclidean coordinates  of the vectors $\vec{a}_j$. These then have a finite number of independent solutions ( up to  overall rotations).  These can be explicitly listed, in principle, for any  $N$, and  given set of masses $\{m_i\}$, using a computer.

In particular, if $N=3$, and all the masses are equal, the positions of the masses form an equilateral triangle, with the velocities directed towards ( or away from) the center of the triangle. This is obvious by symmetry. 

Interestingly, even if the masses are unequal, the positions of masses form an equilateral triangle! This is not {\em immediately } obvious. But may be seen as follows: The forces  on  mass $m_1$  due to the other two masses are easily seen to be of the form $m_2 K \hat{e}_{2,1}$ and $ m_3 K \hat{e}_{3,1}$, where $K$ is some constant, and  $\hat{e}_{3,1}$ and  $ \hat{e}_{2,1}$ are unit vectors in the direction of  the two masses.  Clearly, their sum is  in the direction of the center of mass (see Fig. 1). The rest of the argument is straightforward.

What happens after this collision is not well defined. Physically, it is not possible to approximate masses as point masses, if the distances are very small, and the evolution equation stops being valid. Also, the relativistic corrections will come into play if the velocities are close to the light velocity. 

\begin{figure}
    \centering
    \includegraphics[width=0.65\columnwidth]{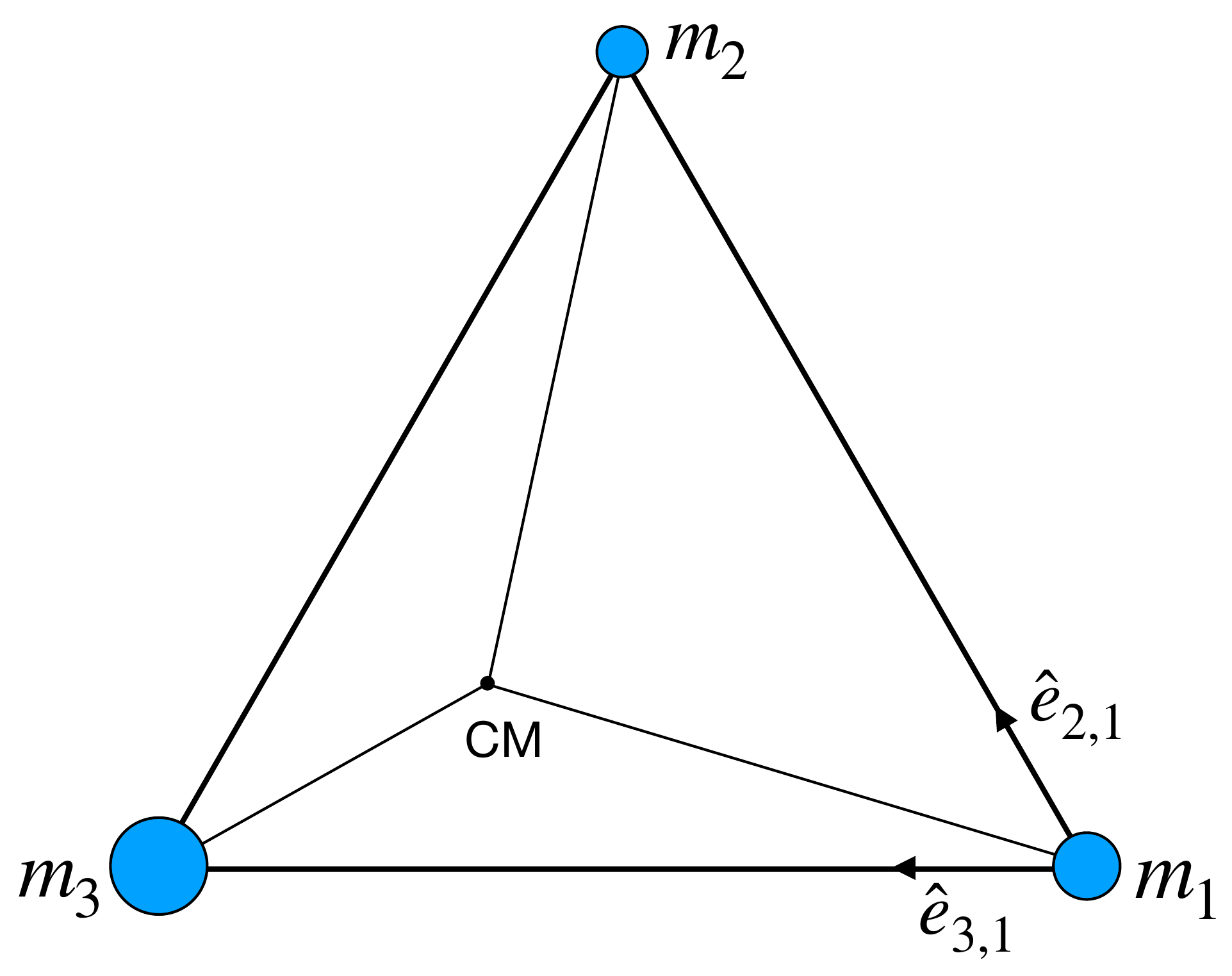}
    \caption{ The scaling solution for three masses in a plane is an equilateral triangle,  even when the masses are unequal. Here $CM$ marks the center of mass, and $\hat{e}_{2,1}$ and $\hat{e}_{3,1}$ are unit vectors along the sides of the triangle, as indicated.  }
    \label{fig:fig1}
\end{figure}

\section{Finite-time singularities}

In physics, we often specify the time evolution of a system, by differential equations. Given, any initial conditions, the equations may be integrated forward ( or backwards) in time, and this specifies the evolution at all times. But this procedure encounters a problem, if the solution develops a singularity, so that, it is not possible to continue the forward integration.  A mathematician would say that solution, as defined by the solution of the differential equation,  {\em does not exist} beyond this time. It is perhaps simpler to say that one has to add to the differential equation a prescription about what happens to the system once the singularity has formed.  

An example will make this clearer. Consider motion in one dimension, with the position of a particle at time specified bu $x(t)$.  We assume that the evolution is governed by the the simple equation $ dx/dt =x^2$, with the initial condition $x(t=0)=1$. It is easy to see that the solution of the equation is $x(t) =1/(1-t)$. The velocity of the particle increases as it moves, and it reaches inifinity at time $t=1$. What happens for $t>1$? The differential equation fails to  specify the evolution beyond this point. We will say that the motion develops  a finite time singularity at $t=1$.  

In this case, common sense will tell us that particle at infinite distance will remain at infinite distance at all subsequent times. This is the extra physical information we have to provide, to answer the question, "What is $x(t)$, for $t>1$". Alternatively, one could specify that as soon as  the particle reaches infinity, it reappears at $x=1$!

As a   slightly more complicated example, consider the motion of a  bouncing ball under gravity.  We will treat the ball as a point mass. say, it is released at rest from a height $H$. When the ball reaches the ground, it bounces back, with the restitution coefficient $\alpha$. This is a standard example taught in secondary schools, and we know that after the rebound, it will reach a height $\alpha^2 H$, and after the second rebound $ \alpha^4 H$ , and so on.  Also, the time between successive bounces decreases exponentially  as the number of bounces increases. In fact, The ball undergoes an infinite number of bounces, in a finite time.  

A person blindly following the forward integration algorithm, would calculate the the velocity when it first hits the ground, then the rebound velocity after first collision with the floor, then the velocity just before it   hits  the floor the second time, then the velocity just after the second rebound, and so on.  Clearly, this process does not terminate, and the equations of evolution and the rules of inelastic collisions at the floor,  by themselves, do not allow us to tell what  happens after infinite number of collisions.  Each collision is a singularity in the evolution equations (velocity is discontinuous).  The time when the ball eventually comes  to rest is  an accumulation point of such discontinuities. This is  a  qualitatively different finite-time singularity. To specify the evolution of  the system past this singularity, some extra specifications  dictated by the physics of the situation have to be added. Here, we will specify that   after all the collisions are over, the ball remains on the floor at all subsequent times.  

Or, as a less contrived example, consider the evolution of  gas cloud, undergoing  gravitational collapse. We  may describe the evolution in terms of partial  differential equations for the density field, velocity field and the temperature of the gas, and the gravitational field. If the evolution leads to a black hole being formed at a finite time, this will be an example of a finite time singularity. To specify further evolution, we have to add to the differential equations what happens to gas clouds near a black hole. [ If the problem was  about the cosmological  Big Crunch, then it is not even clear that "time" has a meaning after the crunch has happened!]

In the following, we will discuss the problem of finite time singularities, where it turns out that all the masses go to infinite distances in a finite time. 

\section{A test mass in the time-varying field of two gravitationally  bound masses}

\begin{figure}[h]
    \centering
    \includegraphics[width=0.65\columnwidth]{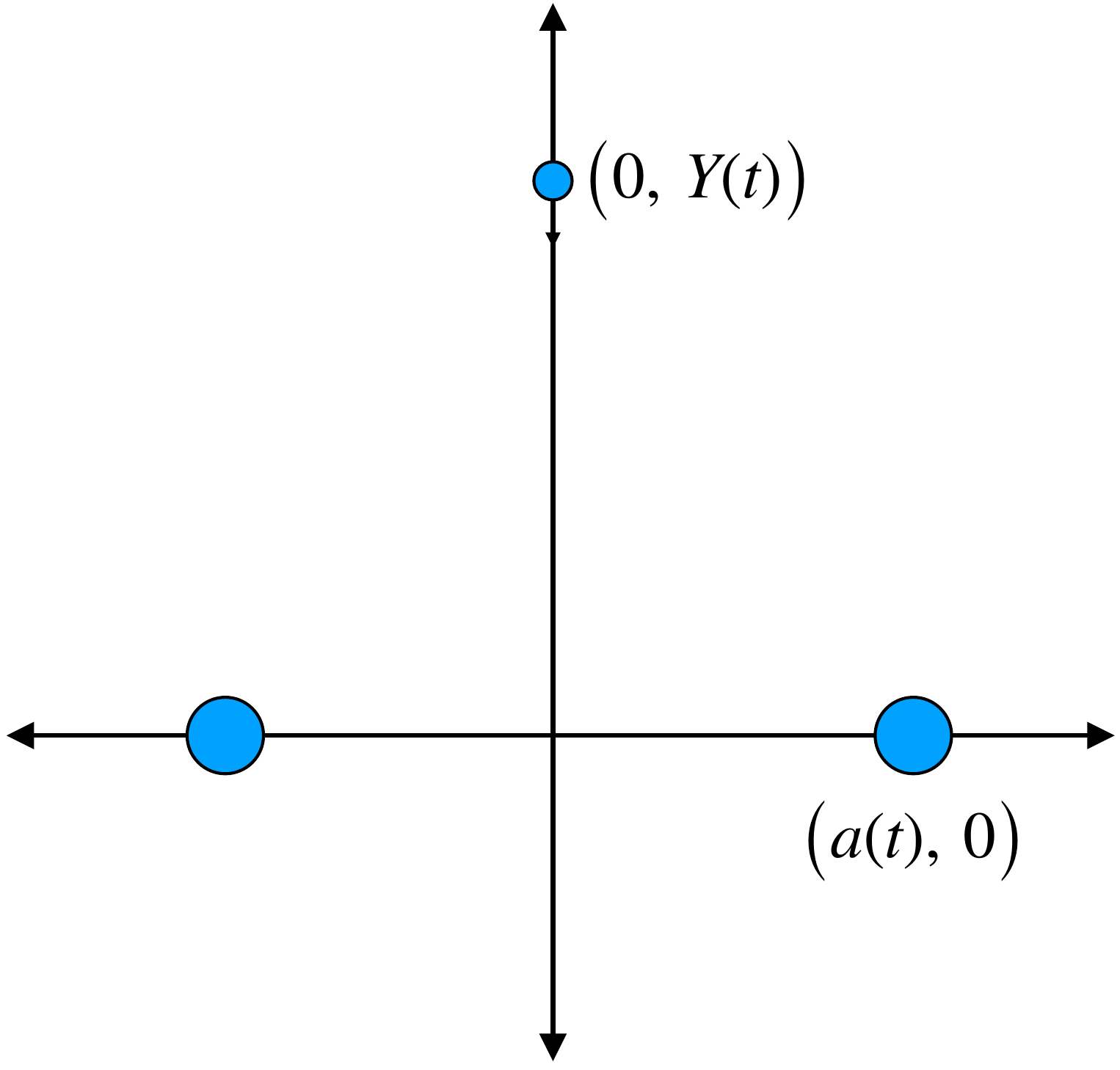}
    \caption{The figure shows  two heavy masses that move along the x-axis, and a  light test mass that moves along the $y$-axis, in the time-dependent gravitational field generated by the two masses.}
    \label{fig:fig2}
\end{figure}

We start by discussing the problem in a simplified setting. We consider two equal masses $M$ that are constrained to move on a straight line, in the center of mass frame. We denote  the positions of the masses  by  
$+a(t)$ and $-a(t)$. Between rebounds, the function $a(t)$ satisfies the equation
$$  \frac{d^2}{dt^2}  a(t) = - \frac{GM}{4   a(t)^{2}}. $$
This is, of course, just  the degenerate  limiting case of the two-body problem, where the ellipse is reduced to a line. It was realized that in this case, one can smooth out the singularities of the two-body collisions, by introducing a  new time-parameter $\tau$, related to $t$ by a non-linear transformation
\begin{equation}
    t = \tau - \sin \tau.
\end{equation}
It is easily verified that the time dependence of position $a(t)$ has a much simpler dependence in terms of this parameter
\begin{equation}
 a(\tau) = a_0 [ 1 - \cos(\tau) ].
\end{equation}
\begin{figure*}
    \centering
    \includegraphics[width=1.9\columnwidth]{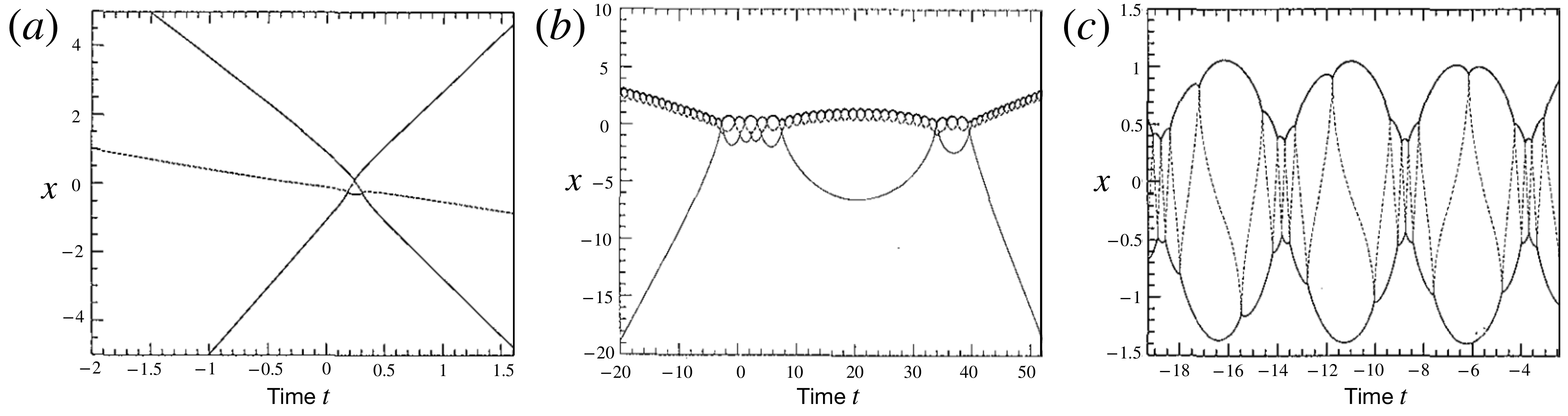}
    \caption{Space-time plot of the motion of a system of three masses on  a line  in the center of mass frame. The x-axis is time, and the $y$-axis is the spatial coordinate. The panels correspond to the three possible scenarios, discussed in the text. Figure adapted from \cite{hietarinta}.}
    \label{fig:fig3}
\end{figure*}
Now, we introduce a test particle of negligible mass $m$ that moves along the $y$-axis, in the time-dependent gravitational field provided by the larger masses. In the limit of small $m$,  this small mass has  negligible effect on the motion of the larger masses. If $y$-coordinate of the position of the test mass is $Y(t)$,  then $Y(t)$ satisfies the differential equation
\begin{equation}
 \frac{d^2}{dt^2} Y(t) = -2 G M \frac {Y(t)}{ [a^2(t) + Y^2(t)]^{3/2}}. 
\end{equation}

We release this test mass from rest at some time-parameter value $\tau_0$ from the position $(0, R_0)$, with $R_0 \gg a_0$. Since we expect that the motion may show strong dependence on $R_0$, we will take $R_0$ to be not exactly known, but with limited precision: All positions in the range $[ R_0- \Delta R, R_0+\Delta R]$ are assumed equally likely, with $\Delta R \ll R_0$. 

Some features of the qualitative behavior of $Y(t)$ are easy to see: It would be attracted towards the larger masses, and move towards the origin. Initially, its velocity is small, but increases as $Y(t)$ decreases, and eventually it crosses the $X$-axis. After crossing the x-axis, it will loss kinetic energy, as it goes away from the attractive masses. Two cases can arise : eventually it reaches a finite minimum $Y(t)$, and then turns back, or it may have acquired  enough kinetic energy to keep going to arbitrarily large  negative values. If the large masses were coming together when the test mass crosses the origin say at time $t_1$, then the attractive force at time $t_1 - \epsilon$ is less than the attractive force at time $t_1 +\epsilon$, and the sum of these is a decrease in the kinetic of the particle. In the opposite case, the kinetic energy of the test particle will increase.

There is no limit to the amount of energy transfer.  If the large masses are at a distance $\delta x$ when the test mass crosses the origin, then the acceleration undergone by the light mass is of order $\frac{1}{(\delta x)^2}$, for a duration that is of order $(\delta x)^{3/2}$, hence the extra velocity acquired is of order $(\delta x)^{-1/2}$. Thus the increase in kinetic energy is of order $(1/\delta x)$. Also, the probability that particle crosses when $\delta x$ is  small is of order$(\delta x)^{3/2}$. The conclusion of this power -counting argument is that in letting the particle drop from a large distance $R_0$, with some uncertainty in the initial position, the excess kinetic energy $\Delta E$ acquired by the particle has a power-law tail: the probability of energy gain $ \geq \Delta E$ varying as  $(\Delta E)^{-3/2}$. 

So, with a significant probability, the particle will go off to $-\infty$ after crossing the $x$-axis. If it  does not get enough energy to leave, it will return to the origin, and depending on the phase of the binary motion at that time, again pick up a large amount of kinetic energy, and go off to infinity,  with a  non-zero velocity in the positive direction. If, it does not get enough energy in the second crossing of the x-axis, there is still possibility at the next  time. Thus, for a particle starting at very large $R_0$, the probability that  goes off to infinity after less than $n$  crossings tends to $1$ for large $n$.  

We started by assuming that $R_0 \gg a_0$. If this is not so, one can get more complicated motions, including periodic orbits, and aperiodic bounded motion.  One can see nice phase space pictures of different types of motion possible in \cite{hietarinta}.

What we learn from this exercise is that when a test mass has a close encounter with a massive binary, there is a finite probability that it will get an increased kinetic energy after the encounter. If we want think of the binary as a single tightly bound object, then collision of the third body with this is like an inelastic collision, {\em  but with a  varying coefficient of restitution.} This coefficient is not of constant value, and its value depends on the details of the encounter, but can be greater than $1$, with a finite probability.

If we now consider the motion of three masses on a line, where the order of masses cant change, then three possible scenarios are possible, shown in Fig.3.\\
(a) Eventually all particles escape to infinity with finite limiting velocities  $V_1^* < V_2^* < V_3^*$.\\
(b) system breaks into two clusters: one single mass and a bound binary, and these have a finite relative velocity with respect to each other. \\
(c) The all the masses remain in a bounded region of space, and execute periodic, or quasi-periodic motion.\\

\section{Four particles on a line}

We are now in a position to describe  the unexpected behavior  we stated in the introduction. We consider four masses, $m_1, m_2,m_3$ and $m_4$, constrained to move along the x-axis under mutual gravitational interactions.   If any two masses come very close to each other, the attractive force becomes stronger, and there is a binary collision.  At this point, the equations of evolutions are ill behaved, and there is a "collisional singularity". We take care of this singularity by specifying that  after the collision, the masses rebound, and cannot cross. If two masses collide at time $t^*$, we assume that the velocities at time $t^*+ \epsilon$ are negative of the velocities at time $t^* -\epsilon$, in the center of mass frame of the two masses. This defines how to continue the evolution, after the binary collision. We will also not discuss the case where three bodies come to the same point at the same time, as they correspond to a very unlikely  initial conditions. 

We will work in the inertial frame where the center of mass of the four masses is at the origin at all times. As the masses can not cross, we will always have the order of masses preserved, with their positions satisfying $x_1 \leq x_2 \leq x_3 \leq x_4$ at all times.  We will take the mass  $m_2 < m_1, m_3,m_4$, and  take the masses $m_3$ and $m_4$ to a tightly bound binary, with their distance $|x_3 - x_4|$ much less than other distances. 

As the time evolves, the system undergoes a sequence of binary collisions. It was shown by  Saari and Xia, and later discussed by Mather and McGehee, that the qualitative result can be described as follows:  the mass $m_2$  keeps colliding  with masses  $m_1$ and $m_3$ alternately.  The mass $m_3$ undergoes several collisions with mass $m_4$ between two consecutive collisions with mass $m_2$. 

We have argued it is plausible, and  one can show, that the overall effect of the collision of mass $m_2$ with the tight binary  $m_3$-$m_4$ is  an inelastic collision, with coefficient of restitution greater than 1. Thus, after rebounce from mass $m_3$, the kinetic energy of mass $m_2$ is increased. On collision with $m_1$, some of this energy is transferred to $m_1$, but  still remains greater that the earlier value. Then after next bounce from the binary, the mass $m_2$ picks up even more energy, again part of it transferred to $m_1$. There is no upper limit to this energy transfer, as the energy comes from the binary system, which becomes more and more tightly bound after each collision. 

The overall description of motion is then as follows, after each collision, the velocity of mass $m_1$ increases on the average by a constant factor greater than $1$, and the velocity of $m_1$ tends to $-\infty$ for large $n$.  The velocity of the the center of mass of the binary $m_3$-$m_4$ tends to $+\infty$, and the maximum separation between the masses $m_3$ and $m_4$ decreases with increasing number of collisions $n$. The velocity of the $m_2$ alternates between positive and negative values, but increases in magnitude. Since the total momentum of system is always zero, all these velocities increase in the same way, say  roughly as $\lambda^n$, where $\lambda$  is the same for all masses.

The distances  between particles also increase exponentially with the collision number $n$, but with a different constant, as  $\mu^n$, with $\mu < \lambda$. Then the time between the $n$-th and $(n+1)$-th collision of a mass decreases with $n$ approximately geometrically, as $(\mu/\lambda)^n$. Then, as in the elastic ball bouncing on a floor, the total time taken for an infinite number of collisions is {\em finite}. 

Thus, we get all the masses tend to infinite distance in a finite time.  This is in spite of the fact that the velocity of each particle remains finite after each collision. It takes an infinite number of collisions to increase the magnitudes of the  velocities to infinity, but these infinite number of collisions occur in  a finite time. 

The arguments given above are qualitative, and heuristic. For more careful, and rigorous derivations, one should consult the literature cited. 

\section{Summary}

In this article, I have tried to discuss the qualitative features of motion of point masses, moving under mutual  gravitational interaction, in Newtonian mechanics.  We discussed the first some features of the scaling solutions, where the finding the solution for the motion  reduces to finding the solution of  a second-order differential equation in one variable. 
 
We then discussed the question of finite-time singularity, when $n$ bodies collide, coming together at the same space-time point.  We discussed a simple case of the test mass in the field of oscillating binary masses to show that the mass can undergo very large accelerations. Finally, we discussed qualitatively the case of four masses on a line, where the third and fourth masses at one end form a tight binary, and the on each collision with this binary, the second mass keeps increasing its kinetic energy.  Then, the kinetic energy of all the masses increases exponentially with the number of collisions, but as the time between collisions decreases exponentially, all masses go to infinity in a finite time, undergoing infinite number of collisions.

It is useful to emphasize here, that while this result is rigorous, its validity depends on the validity of the assumed evolution equations, i.e. in the  the point -mass and non-relativistic approximations.     Also, it happens only for a very special set of initial conditions, which have a very small measure in the phase space.  Even so, I find it interesting, and instructive. I hope that you did, too.

\acknowledgements{I  thank Rajaram Nityananda for discussions and  comments on the draft script, Aanjaneya Kumar for help in preparing the manuscript for submission, and an IISERP student, Pushparaj Chakravarti. Extended discussions with Pushparaj  convinced  me of the usefulness of an expository article on this topic.}

\end{document}